\documentclass[letter]{aa} 

\usepackage{txfonts}
\usepackage{graphicx}   
\usepackage{amsmath}    
\usepackage[utf8]{inputenc} 
\usepackage{array} 
\usepackage{multicol,rotating} 
\usepackage{longtable} 
\usepackage[section]{placeins}
\usepackage{lscape}
\usepackage{subcaption}
\usepackage{multirow} 

\begin{document} 

\title{Size, albedo and rotational period of the Hayabusa2\# target (98943) 2001 CC21} 

  \author{S. Fornasier
          \inst{1,2}
          \and
          E. Dotto\inst{3} \and
          P. Panuzzo\inst{4} \and
          M. Delbo\inst{5,6} \and
          I. Belskaya\inst{7,1} \and
          Y. Krugly\inst{7} \and
          R. Inasaridze\inst{8,9} \and
          M. A. Barucci\inst{1} \and
          D. Perna\inst{3} \and
          J. Brucato\inst{10} \and
          M. Birlan\inst{11,12}
           }

   \institute{LESIA, Universit\'e Paris Cit\'e, Observatoire de Paris, Universit\'e PSL, CNRS, Sorbonne Universit\'e, 5 place Jules Janssen, 92195 Meudon, France
              \email{sonia.fornasier@obspm.fr}
         \and
           Institut Universitaire de France (IUF), 1 rue Descartes, 75231 PARIS CEDEX 05  
        \and  
        INAF-Osservatorio Astronomico di Roma, 00078, Monte Porzio Catone (Roma), Via Frascati 33, Italy
        \and
        GEPI, Observatoire de Paris, Université PSL, CNRS, Place Jules Janssen, 92195, Meudon, France
        \and
        Université Côte d'Azur, CNRS-Lagrange, Observatoire de la Côte d'Azur, CS 34229, 06304, Nice Cedex 4, France
        \and
        11 School of Physics and Astronomy, University of Leicester, UK
        \and 
        Institute of Astronomy, V.N. Karazin Kharkiv National University, 4 Svobody Sq., Kharkiv 61022, Ukraine 
        \and
        E. Kharadze Georgian National Astrophysical Observatory, Abastumani, Georgia
        \and 
        Samtskhe-Javakheti State University, Akhaltsikhe, Georgia
        \and  INAF-Astrophysical Observatory of Arcetri, Largo E. Fermi 5 50125 Firenze, Italy
        \and 
          IMCCE, Observatoire de Paris, CNRS, PSL Research University, 77 av. Denfert Rochereau, 75014, Paris Cedex, France
        \and
        Astronomical Institute of the Romanian Academy, 5 Cutitul de Argint, 040557, sector 4, Bucharest, Romania            
             }

   \date{Received on 2024}

 
  \abstract
   {}
   {This study aims to determine the size, albedo and rotational period of (98943) 2001 CC21, target of the Hayabusa2 extended mission, using thermal data from the {\it Spitzer} Space telescope and ground based observations.}
   {The {\it Spitzer} data were acquired with the Infrared Spectrograph in the 6-38 $\mu$m range, reduced using the {\it Spitzer} pipeline and modeled with the Near Earth Asteroid Thermal Modeling to determine the asteroid size and albedo.  The absolute magnitude and the rotational period were determined thanks to new observations carried out at the 3.5m New Technology Telescope, at the 1.2m Observatoire de Haute Provence, and at the 0.7m Abastumani telescope. Three complete lightcurves were obtained in 2023-2024 at the last mentioned telescope.}
   {We determine an absolute magnitude of H=18.94$\pm$0.05, and a rotational period of 5.02124$\pm$0.00001 hours, with a large lightcurve amplitude of $\sim$ 0.8 mag. at a phase angle of 22$^o$, indicating a very elongated shape  with estimated a/b semiaxis ratio $\geq$  1.7, or a close-contact binary body.  The emissivity of 2001 CC21 is consistent with that of silicates, and its albedo is  21.6$\pm$1.6 \%.  Finally, the spherical-equivalent  diameter of 2001 CC21 is 465$\pm$15 m.}
   {The albedo value and emissivity here determined, coupled with results from polarimetry and spectroscopy from the literature, confirm that 2001 CC21 is an S-complex asteroid, and not a L-type, as previously suggested. The size of 2001 CC21 is less than 500 m, which is smaller than its first size estimation ($\sim$ 700 m). These results are relevant in preparation of the observing strategy of 2001 CC21 by Hayabusa2 extended mission.}


   \maketitle
%

\section{Introduction}

The Japan Aerospace Exploration Agency's (JAXA) Hayabusa2 mission,  following the successful return of the Ryugu samples in December 2020, has been extended to explore two more Near Earth Asteroids (NEAs): (98943) 2001 CC21, which is scheduled for a flyby in 2026, and the fast-spinning 1998 KY26, for a rendez-vous on 2031. The extended mission has been nicknamed Hayabusa2\#, where the  \# character stands for "SHARP" (Small Hazardous Asteroid Reconnaissance Probe). Several observing campaigns of these two targets have been and will be carried out to better understand their physical properties in support of the Hayabusa2\# mission, and in particular to optimize the observing strategy. \\
This paper focuses on 2001 CC21, which is an Apollo NEA. The first spectrum of NEA (98943) 2001 CC21 was obtained during the MITNEOS survey in October 2004, and leaded to an L-type taxonomic classification \citep{Binzel_2004, Binzel_2019}. This type of asteroid should be rich in calcium-aluminum inclusions \citep{Sunshine_2008}, and having a flat spectrum in the 0.75-1 $\mu$m region. However, other spectra obtained in the literature \citep{Lazzarin_2005, Geem_2023} clearly indicate the presence of well-defined absorption features in the 0.9-1 $\mu$m range, and around 2 $\mu$m, associated with the presence of pyroxene. Therefore, 2001 CC21 belongs to the S-complex and has been classified as Sq, Sk or Q type in the Bus-Demeo classification scheme \citep{Demeo_2009}. \\
In addition, the polarimetric properties of 2001 CC21 determined by \cite{Geem_2023} are consistent with those of S-complex asteroids and different from the ones of L-type asteroids. These authors also obtained the first measurements of the target albedo of 23$\pm$4 \% from the polarimetric slope, and estimated a size ranging from  440 m to 530 m. \\
This body is known to have an elongated shape, with a lightcurve amplitude of 0.75-1.1 magnitudes, and a rotational period of 5.0159 h \citep{Warner_2023}, or 5.0247 h\footnote{\url{https://www.asu.cas.cz/~ppravec/newres.txt}}.\\

In this paper we present the results of the analysis of the spectral energy distribution (SED) of 2001 CC21 obtained by the {\it Spitzer} Space Telescope in the 6-38 $\mu$m range which allows us to accurately estimate the size and albedo of the target. 
We also present ground-based observations in photometry performed to determine the rotational period of 2001 CC21, and its absolute magnitude. This last is a basic input parameter of the thermal model. 

\section{Rotational period and photometry}

\begin{figure}
\centering
\includegraphics[width=0.49\textwidth]{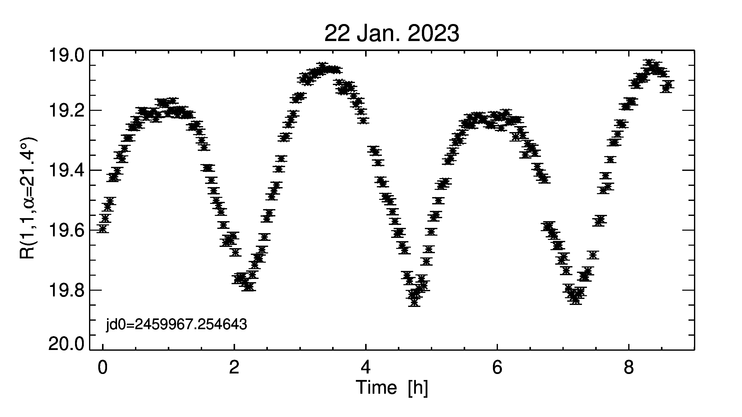}
\includegraphics[width=0.49\textwidth]{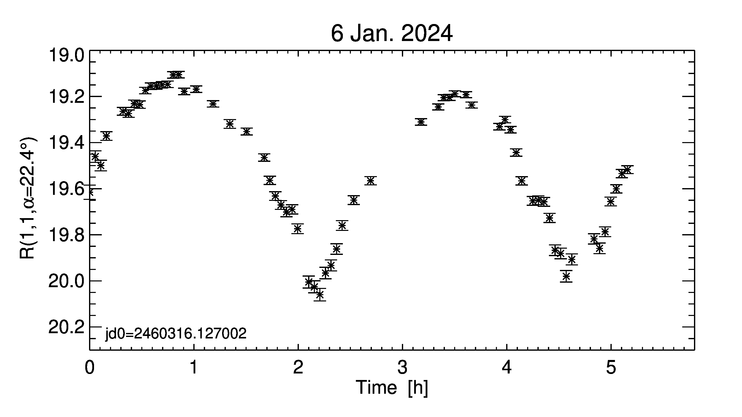}
\includegraphics[width=0.49\textwidth]{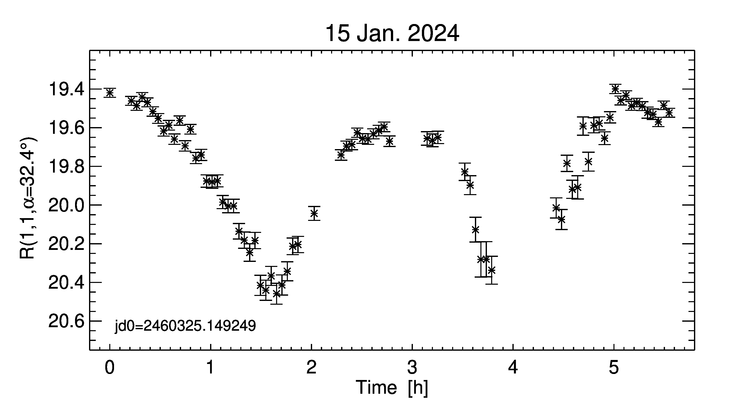}
\caption{(98943) 2001 CC21 lightcurves, in reduced R magnitude,  acquired on January 22, 2023, and on January 6 and 15, 2024 at the 0.7 m Abastumani telescope.}
\label{lc}
\end{figure}

\begin{figure}
\centering
\includegraphics[width=0.47\textwidth]{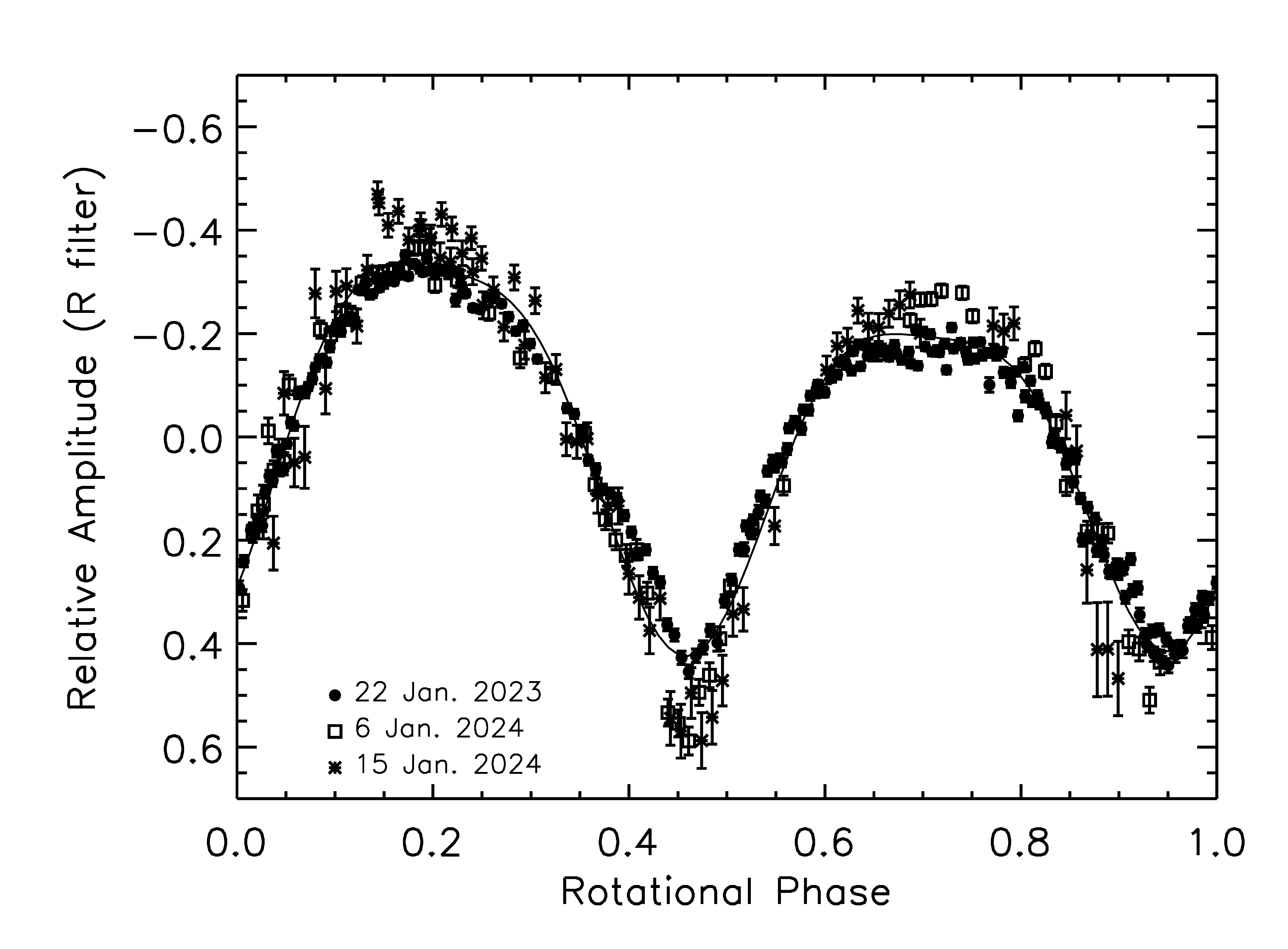}
\caption{Composite lightcurve of 2001 CC21. The zero phase time corresponds to JD 2460100.}
\label{phase_2001CC21together}
\end{figure}
  
\begin{figure}
\centering
\includegraphics[width=0.47\textwidth]{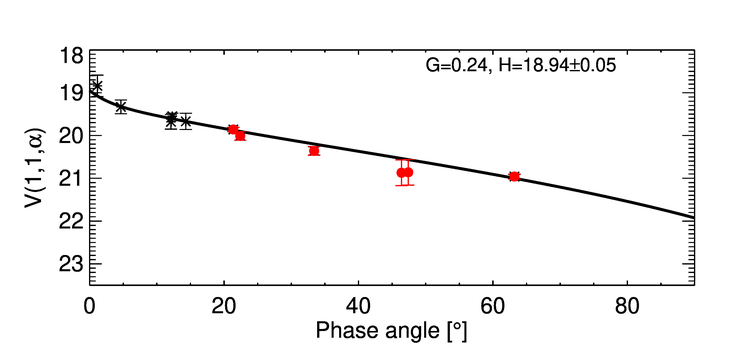}
\caption{HG fit of the photometric data here presented (Table~\ref{tab_phot}, red circles), together with data from astrometry available from the MPC database.}
\label{phase_2001CC21}
\end{figure}

To support the {\it Spitzer} data interpretation, we carried out photometry of 2001 CC21 on October 31, 2005 at the 3.58 m New Technology Telescope (NTT).  We used the EMMI instrument in the V, R, and I filters centered at 543, 641, and 799 nm, coupled with a 2048$\times$4096 CCD, having squared pixel of 15 $\mu$m. Further photometric observations were conducted at the 1.2 m telescope of the Observatoire de Haute Provence (OHP) in November 2022 in the broadband Johnson B, V, R, and Gunn I filters. The data were collected with a 2048$\times$2048 detector in 2$\times$2 binning mode, resulting in a pixel scale of 0.77 arcsec px$^{-1}$. \\
Additionally, 3 lightcurves were obtained in the Johnson-Cousins R filter at 0.7 m Abastumani Miniscus telescope (Georgia) in January 2023 and 2024 (Fig.~\ref{lc}). These observations were obtained with a 2048$\times$2048 CCD in unbinning mode with a pixel scale of 1.3 arcsec px$^{-1}$.  The data were reduced in a standard way \citep{Fornasier_2004}, fluxes were extracted via aperture photometry and absolutely calibrated using different standard stars observed during each nigh \citep{Landolt_1992}.  The observing conditions and results are reported in Table~\ref{tab_phot}. The uncertainties in magnitude and colors take into account both the instrumental and the calibration errors. 

We applied the standard Fourier analysis \citep{Harris_1989} to these lightcurves, obtaining a rotation period of 5.02124$\pm$0.00001 hours. This value is also confirmed when we include in our analysis two lightcurves obtained on January 31 and February 1$^{st}$ 2023 with the 0.3 m telescope (Observatorio Estelia Ladines, Asturias, Spain) and publicly available in the Asteroid Lightcurve Data Exchange Format (ALCDEF) database\footnote{\url{https://alcdef.org/}}. Fig.~\ref{phase_2001CC21together} shows the composite lightcurve obtained by combining the 3 single night lightcurves reported in Fig.~\ref{lc}. \\

The amplitude of the light curve changed from 0.77 in 2023 to 0.86 in 2024 at phase angle about 22$^o$, and increased to 1 mag for observations at phase angle of 33$^o$.  In the literature the highest amplitude of 1.1 mag was measured by Pravec in 2022 at a phase angle of 68$^o$. Considering the lightcurve amplitudes measured in 2002-2024 in the literature and in this paper, and the amplitude-phase angle relationship for S-type asteroids \citep{Zappala_1990}, the 2001 CC21 amplitude reduced to zero phase angle is in the range of 0.36-0.58 mag. Using the largest reduced amplitude, we estimate for 2001 CC21  an elongated shape with an a/b half-axis ratio $\ge$ 1.7. Since the light curves show broad maxima and sharp minima, we can assume a contact binary nature for this asteroid, although it is impossible to distinguish an elongated from a contact binary object based on the light curve shape alone \citep{Harris_2020}.




To compute the absolute magnitude, we used the entire data set presented in this paper, as well as data available in the literature. 
Numerous estimations of the asteroid's magnitudes were made during its astrometric observations in 2001-2024 covering the wide phase angle range from 1 to 96$^{o}$. These data are available in the MPC database\footnote{\url{https://www.minorplanetcenter.net/db_search/show_object?object_id=98943}}. To obtain the phase curve we used magnitude measurements in V and R bands, and we assumed V-R color index of 0.43, which is the average value determined from the NTT and OHP observations (Table~\ref{tab_phot}), to combine all measurements in V band. We calculated the average magnitude from astrometric observations at a given phase angle using typically 5-10 measurements, and we took the standard deviation as the magnitude error. \\
We first applied the HG system \citep{Harris_1989}, assuming a G value of 0.24, which is the standard value for S-type asteroids \citep{Warner_2009, Pravec_2012}, including the data presented in Table~\ref{tab_phot} and some data from the literature covering the smaller phase angle values. The mean magnitudes from the lightcurve studies and the NTT observations taken a few weeks before the {\it Spitzer} one (which turned out to be taken close to the lightcurve mean, as shown in section 3) were given a higher weight in the fit.  The derived H value (in the V filter) is H = 18.94$\pm$0.05 (Fig.~\ref{phase_2001CC21}). \\
We also checked the consistency of this result by analyzing all the data here presented and those available from the literature with the HG1G2 system \citep{Muinonen_2010}. We obtained H=18.92$\pm$0.20, G1=0.2588, and G2=0.3721 (Fig.~\ref{hg1g2}). The large uncertainty in the absolute magnitude is due to the wide amplitude of the 2001 CC21 lightcurve, and to the large error bars of the astrometric observations.



\section{{\it Spitzer} observations and data reduction}

\begin{table}
\small{
\begin{center}
\caption{Observational details of 2001 CC21 with {\it Spitzer}/IRS ({\it Seg.} represents the segments name). } 
\label{tab_obs}
\begin{tabular}{c|c|c|c|c} \hline
 \textbf{Date} & \textbf{UT$_{start}$} & \textbf{T$_{ramp}$ (s)} & \textbf{Total T$_{exp}$ (s) } & \textbf{Seg.}  \\  \hline 
 2005-11-20&10:17:06.340 & 14.68 &  88.08    & SL2  \\  
 2005-11-20&10:19:37.039 & 14.68 &  88.08    & SL2   \\ 
 2005-11-20&10:22:10.339 & 14.68 &  58.72    & SL1    \\ 
 2005-11-20&10:23:57.042 & 14.68 &  58.72    & SL1   \\ 
 2005-11-20&10:26:10.245 & 121.9 &  1218.96  & LL2    \\ 
 2005-11-20&10:51:21.747 & 121.9 &  1218.96  & LL2   \\ 
 2005-11-20&11:16:33.242 & 121.9 &  2437.92  & LL1   \\ 
 2005-11-20&12:06:47.737 & 121.9 &  2437.92  & LL1    \\ 
\hline
\end{tabular}
\tablefoot{The observations were carried out at heliocentric and target-{\it Spitzer} distances of 1.16824 au and  0.29372 au, respectively, and at a phase angle of 52.6$^{\circ}$.}
\end{center}
}
\end{table}

NEA (98943) 2001 CC21 was observed on November 20, 2005 from 10:17 to 12:26 UT with the Infrared Spectrograph \citep[IRS,][]{Werner_2004} onboard the {\it Spitzer} Space Telescope. Data were acquired in the low resolution mode ($R = \lambda/\Delta\lambda \sim$ 64-128) covering the 5.2--38 $\mu$m range using the four IRS long slit segments:
the 1$^{st}$ and 2$^{nd}$ orders of the short wavelength, covering the 7.4--14.2 $\mu$m (SL1) and 5.2--8.5 $\mu$m (SL2) wavelength ranges, respectively; and the 1$^{st}$ and 2$^{nd}$ orders of the long wavelength segment, covering the 19.5--38.0 $\mu$m (LL1) and the  14.0 --21.5 $\mu$m ranges (LL2), respectively. The individual ramp times and the total exposure time are reported in Table~\ref{tab_obs} together with the observing conditions of 2001 CC21.\\
The data were reduced starting from the basic calibrated data generated by the {\it Spitzer} Space Center automated pipeline, which are corrected for flat fielding, dark current, stray light, and cosmic rays, and absolutely calibrated in flux \citep{Houck_2004}.   
The sky background was removed by differencing two consecutive images taken at different nodding positions for each spectral segment. \\
Finally, we used the {\it Spitzer} IRS Custom Extraction (SPICE) software to extract the one-dimensional spectra for each of the four IRS segments \citep{Houck_2004}, following the methodology and the steps described in \cite{Lamy_2008}, and \cite{Barucci_2008}. \\
The estimated photometric uncertainties are typically $\pm$2\%, but can reach 5\% at the edges of the different segments\footnote{\url{https://irsa.ipac.caltech.edu/data/SPITZER/docs/irs/irsinstrumenthandbook/IRS_Instrument_Handbook.pdf}}.

\begin{figure}
\centering
\includegraphics[width=0.49\textwidth]{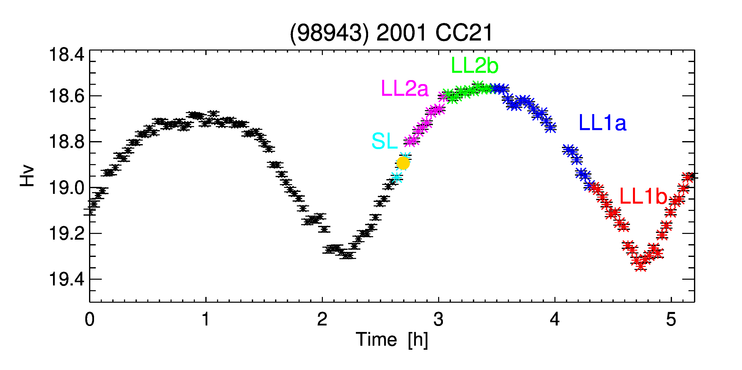}
\caption{The lightcurve of 2001 CC21, in absolute magnitude, obtained on January 22, 2023, with the 0.7 m Abastumani telescope. The approximate estimated positions of the {\it Spitzer} observations are shown in color. The yellow circle represents the V magnitude from the NTT telescope observations acquired three weeks prior to the {\it Spitzer} ones.}  
\label{lc_2001CC21}
\end{figure}

For each IRS segment we have two nodding positions, so two independent spectra to average. The observations of 2001 CC21 cover about half of the rotation period, therefore spanning the full lightcurve amplitude. In particular, flux differences between the first (named a) and the second (named b) nodding positions of the LL1 and LL2 segments (Fig.~\ref{4plots_2001CC21}) are clearly visible, while there are no flux variations for the a and b positions of the SL1 and SL2 segments due to the relatively short exposure time.\\
Therefore, special care was taken in combining the 4 segments according to the lightcurve: the LL2a and LL1a spectra are at similar  flux level, while the LL2b and LL1b spectra have a lower and a higher flux corresponding to a maximum and a minimum of the visible lightcurve, respectively. We present in Fig.~\ref{lc_2001CC21} the January 2023 lightcurve with superposed, in color, the estimated positions of the 4 IRS segments observations. These positions were estimated considering both the time interval between each {\it Spitzer} spectrum and the SL2a start time, and the differences between the thermal fluxes of two adjacent spectra, converted in $\Delta$H$_v$. To combine the SED of the 4 segments we proceed as follows:
a) we used the mean spectra for each of the 4 segments; b) the SL1 and SL2 segments were naturally connected, because the flux was at the same level; c) we shifted the LL2 mean spectrum to match the latest wavelengths of the SL1 one; d) we shifted the LL1 mean spectrum to properly connect it to the shifted LL2 mean spectrum.  

Given the previously determined rotation period, the NTT observations, taken 3 weeks before the {\it Spitzer} data, fall at a rotational phase of +0.0149 after the start of the {\it Spitzer} data, corresponding to a difference of 4.52 minutes.  The absolute magnitude of the NTT observation (yellow symbol in Fig.~\ref{lc_2001CC21}) is 18.89$\pm$0.05, assuming G=0.24.\\
Considering that we have merged the {\it Spitzer} data with respect to the SL segments, and that the $\sim$ 2.5 h duration of the observations covers the full amplitude of the lightcurve, we decided to take as absolute magnitude value for the thermal modeling the one corresponding to the mean value of the lightcurve obtained with the HG system, i.e. 18.94$\pm$0.05. This value is also very close to the one determined from the NTT observations.

 


\section{Size, albedo, and emissivity}

To determine the size (diameter $D$) and albedo ($p_v$) of 2001 CC21, we used the Near-Earth Thermal Model (NEATM, \cite{Harris_1998}, for details see the appendix, section~\ref{text_model}). 

\begin{figure}
\centering
\includegraphics[width=0.45\textwidth]{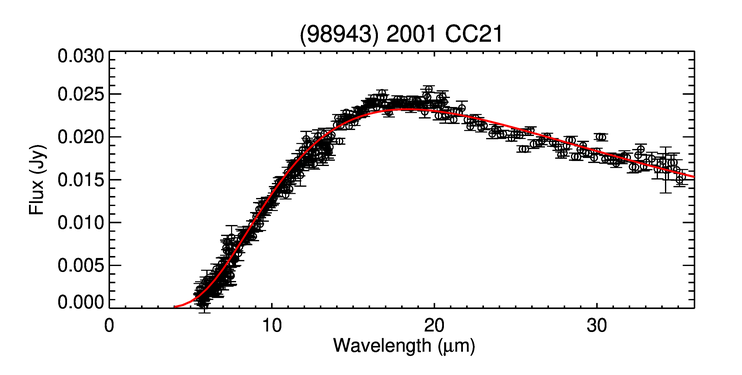}
\caption{Spectral energy distribution of (98943) 2001 CC21  with superimposed the best NEATM fit (red line).}
\label{neatm}
\end{figure}

Using H$_v$=18.94$\pm$0.05, the NEATM best fit results for the 2001 CC21 spectral energy distribution are as follows: diameter = 465$\pm$15 m, albedo p$_V$= 21.6$\pm$1.6 \%, and beaming parameter $\eta$ = 1.847$\pm$0.034. The uncertainties take into account the flux errors, and the {\it Spitzer} uncertainties in the absolute flux values (we use a conservative value of 5\%)\footnote{\url{https://irsa.ipac.caltech.edu/data/SPITZER/docs/irs/irsinstrumenthandbook/IRS_Instrument_Handbook.pdf}}. The derived sub-solar temperature (T$_{SS}$) is 312 K. The observed SED and the NEATM model that best fits the data are shown in Fig.~\ref{neatm}. The derived diameter is much lower than the 700 m originally evaluated \citep{Hirabayashi_2021}, and it falls in the size estimate (440-530 m) from polarimetric observations \citep{Geem_2023}. These authors also found an albedo value of 23$\pm$4\%, very close to the one determined here. Additional thermal data were obtained by NEOWISE and led to a preliminary 2001 CC21 size assessment of 329$^{+78}_{-41}$ m, a value smaller than the one we found, and with much larger uncertainties \citep{Wright_2023}. Finally, on March 5, 2023, a stellar occultation of 2001 CC21 was observed\footnote{\url{http://www.unmannedspaceflight.com/index.php?showtopic=4920&pid=262173&st=975&#entry262173}}, giving an  occultation chord of 449$\pm$12 m. Therefore, the minimal projected dimension of this NEA should be 449 m. With the rotational period estimated here, this event falls at 1.46 h after the zero reference time  in the lightcurve taken on January 23, 2023 (Fig.~\ref{lc}). The corresponding magnitude is in between the secondary maximum and the lightcurve mean, so the projection of this chord should roughly correspond to  the equivalent spherical shape diameter of 2001 CC21.\\
The $\eta$ value of 2001 CC21 is relatively high compared to the 0.756 value used in the standard thermal model of main belt asteroids \citep{Lebofsky_1986}, but rather common for NEAs observed at relatively high phase angle \citep{Mainzer_2014}. Indeed, the {\it Spitzer} observations were made at phase angle of 52.6$^o$, and a correlation between $\eta$ and $\alpha$ has been reported in the literature \citep{Delbo_2004, Wolters_2008, Mainzer_2011, Mainzer_2014}. In fact, in the NEATM model, the beaming parameter increases at higher phase angles to compensate for the loss of the thermal flux, which is normally sent mostly in the sunward direction. 
\cite{Wolters_2008} reported a possible anti correlation between $\eta$ and $p_v$ (their Fig. 9g) for S- and Q-type near-Earth asteroids. 
The $\eta$ and $p_v$ values we found for 2001 CC21 fit very well with the line showing the anti correlation mentioned above \citep{Wolters_2008}. The 2001 CC21 albedo and beaming parameter values are close, within uncertainties, to those found for the binary NEAs (5381) Sekhmet and 2003 YT1 \citep{Delbo_2011}, even though these bodies are at least 3 times larger than the Hayabusa2\# target and rotate faster (rotation period of 2.3-2.5 h). Considering also the shape of the visible lightcurve, this may support the hypothesis that 2001 CC21 is a contact binary, even though this cannot be conclusively deduced by the thermal properties. More generally, Apollo NEAs have $\eta$ value peaking at about 1.3, but with large distribution including values in the 0.8-3 range \citep{Mainzer_2012}. These last authors reported larger $\eta$ values for the Aten population than for the Apollo one, but stress that this difference may be simply correlated with the larger phase angle conditions of Aten versus Apollo observations.

As the shape and spin state of 2001 CC21 remain undetermined, thermophysical models (TPM) cannot be readily employed. The principal advantage of utilizing a TPM for the analysis of thermal IR data is its capability to ascertain the surface thermal inertia, which is a parameter that quantifies the resistance to temperature changes.  Thermal inertia can be used to constrain the nature of the surface  \citep [see][for a review]{Delbo_2015}. Nevertheless, thermal inertia can still be estimated from the beaming parameter value and knowledge of the rotation period and albedo of an asteroid using the method proposed by \cite{Harris_2016}, although with less accuracy compared to using the TPM. The application of this method to 2001 CC21 data results in a thermal inertia of 398$\pm$8 J~m$^{-2}$~s$^{-0.5}$~K$^{-1}$, using Eq. 2 from \cite{Harris_2016}, and assuming the asteroid to have a 90$^o$ aspect angle, which is the most likely solution \citep{Fatka_2023}. In fact, it has been observed that most small NEAs are in YORP end states with an obliquity of 90 degrees. However, the thermal inertia estimated by this method is a function of the unknown aspect angle of 2001 CC21: the lower the aspect angle, the higher the estimated thermal inertia value would be according to the curve shown in Fig.~\ref{Tinertia}. Therefore, we determine here a lower limit of the thermal inertia for 2001 CC21, which should be strictly taken as the nominal value minus 3 $\sigma$ of the aforementioned uncertainty. This corresponds to a $\sim$ 370 J~m$^{-2}$~s$^{-0.5}$~K$^{-1}$ lower limit. However, we stress that this value is a preliminary estimation derived from the $\eta$ value and using the NEATM model, which assumes a spherical shape, that is obviously not the one of 2001 CC21.
A thermal inertia $>$ 370  J~m$^{-2}$~s$^{-0.5}$~K$^{-1}$. is typical of NEAs of the size of 2001 CC21 \citep[see Fig. 7 in][]{Novakovic_2024}. Based on the assumption of an ordinary chondrite composition and thermal properties, it can be reasonably inferred that the lower limit of the thermal inertia estimated here is related to a surface comprising a mixture of regolith and exposed rocks.


\begin{figure}
\centering
\includegraphics[width=0.45\textwidth]{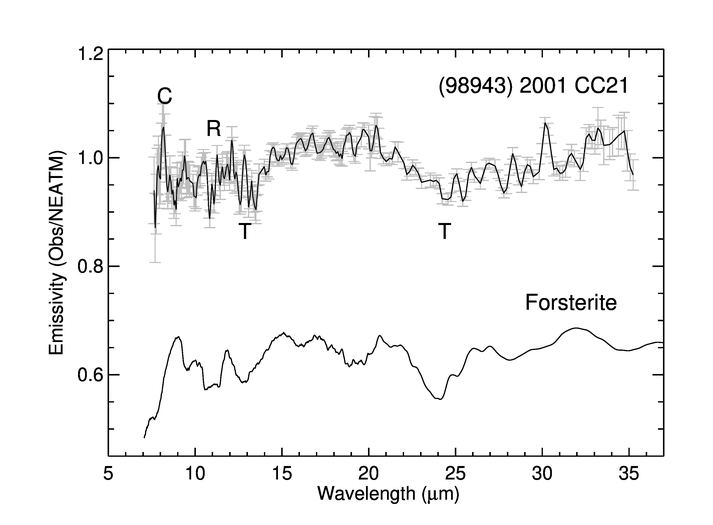}
\caption{Spectral emissivity of (98943) 2001 CC21, with the associated uncertainties (gray symbols), compared with the laboratory emissivity of forsterite. Christiansen, Reststrahlen and Transparency features are noted as C, R and T, respectively.}
\label{emi_2001CC21}
\end{figure}

Figure~\ref{emi_2001CC21} shows the emissivity we obtained for 2001 CC21, as the ratio of the {\it Spitzer} spectrum to the NEATM best fit model (Fig.~\ref{neatm}). The main features observable in this range are the Christiansen peak, reststrahlen, and transparency features.
The Christiansen peak is associated with the principal molecular vibration band and for silicates occurs between 7.5 and 9.5 $\mu$m. In the 2001 CC21 emissivity spectrum, it appears at about 8 $\mu$m. The reststrahlen features, due to the vibrational modes of molecular complexes, have a lower contrast for smaller grain sizes and appear as a plateau between 9 and 12 $\mu$m. Transparency features occur in the spectral region where the absorption coefficient decreases and grains become more transparent. For 2001 CC21 they occur at about 12 $\mu$m, with a secondary feature at $\sim$ 24 $\mu$m.
In order to interpret these features above the thermal continuum in terms of the surface composition of 2001 CC21, we compared its emissivity with that of minerals and meteorites acquired by our group and published by \cite{Dotto_2004, Dotto_2000} and \cite{Barucci_2002}, or available in the literature (\cite{Salisbury_1991b, Salisbury_1991a}, ASTER spectral library\footnote{\url{http://speclib.jpl.nasa.gov}}, and RELAB database\footnote{\url{https://sites.brown.edu/relab/relab-spectral-database}}).
On the basis of this analysis, the emissivity spectrum of our target seems to be in good agreement with the laboratory emissivity spectrum of forsterite, a Mg$_2$SiO$_4$ silicate of the olivine group. 2001 CC21 confirms also in this wavelength range its silicatic nature.

\section{Conclusion}

Utilizing thermal data from the {\it Spitzer} Space Telescope and novel ground-based observations, we have derived the most precise estimation to date of the diameter (465$\pm$15 m), albedo (21.6$\pm$1.6\%), and rotational period (5.02124$\pm$0.00001 hours) of (98943) 2001 CC21. Its thermal inertia, estimated from the beaming parameter, is  $>$ 370 J~m$^{-2}$~s$^{-0.5}$~K$^{-1}$. The albedo and emissivity determined here, along with results derived from polarimetry and spectroscopy in the literature, confirm that the Hayabusa2\# extended mission target is an S-complex asteroid. The results presented here indicate that 2001 CC21 is an elongated asteroid with an a/b ratio of $\ge$ 1.7, or a contact binary.


\begin{acknowledgements}
The authors thank the financial support by Centre national d'études spatiales (CNES), and by the Italian Space Agency (grants ASI/INAF n. 2018-27-HH.0 and 2022-12-HH.0). IB thanks the French PAUSE program, which provides support to scientists at risk. This work utilized data obtained from the Asteroid Lightcurve Data Exchange Format (ALCDEF) database, which is supported by funding from NASA grant 80NSSC18K0851. The lightcurve observations obtained by E.F Mananes are appreciated. 
\end{acknowledgements}

%
%




\newpage

\begin{appendix}
\onecolumn
\section{Supplementary material: Tables and Figures}
\begin{table*}[th]
\caption{Observing circumstances, magnitude and colors of (98943) 2001 CC21 from the 3.5 m NTT, 1.2 m OHP, and 0.7 m Abastumani telescopes. $^*$: plus one day.}
\label{tab_phot}
{\small
\centering
  \begin{tabular}{ccccccccccc} \hline \hline
  Telescope& date     & UT     & r (AU)  & $\Delta$ (AU) & $\alpha$ ($^o$) & V   &  R & B-V & V-R              &   V-I           \\  \hline
3.5m NTT & 31-10-2005 & 03:18--03:26 & 1.108  &  0.418 &   63.2 & 19.27$\pm$0.02   &  &   &   0.44$\pm$0.02   &  0.85$\pm$0.03   \\   
1.2m OHP & 26-11-2022 & 01:36--02:30    & 1.253 & 0.496 &  47.4 &  19.83$\pm$0.03  & & 0.95$\pm$0.06  &   0.39$\pm$0.05	  & \\			
1.2m OHP & 30-11-2022 & 03:25--04:31    & 1.250 & 0.469 & 46.4 &   19.71$\pm$0.03  & & 0.85$\pm$0.04  &   0.47$\pm$0.03    & 0.82$\pm$0.09  \\	
0.7m Abas. & 22-01-2023 & 18:08--02:44$^*$ & 1.149 & 0.179 & 21.4 &  & 15.99$\pm$0.03 &  &  & \\ 
0.7m Abas. & 06-01-2024 & 15:05--20:25  & 1.226 & 0.268 & 22.4 &  & 17.17$\pm$0.05 &  &  & \\ 
0.7m Abas. & 15-01-2024 & 15:37--21:10  & 1.209 & 0.286 & 33.4 &  & 17.63$\pm$0.05 &  &  & \\ 
\hline
\end{tabular}
}
\end{table*}


\begin{figure}[h]
\centering
\includegraphics[width=0.6\textwidth]{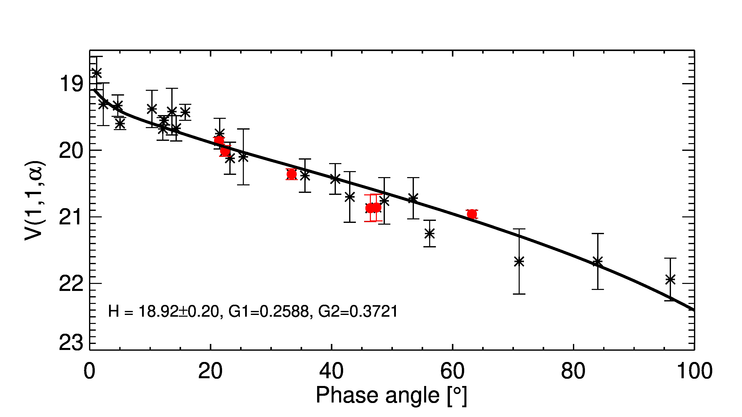}
\caption{HG1G2 fit of the photometric data here presented (Table~\ref{tab_phot}, red circles), together with data from astrometry available from the MPC database.}
\label{hg1g2}
\end{figure}

\begin{figure*}[h]
\centering
\includegraphics[width=0.99\textwidth]{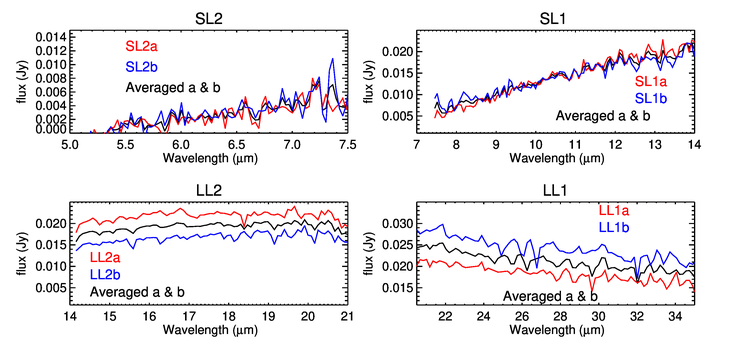}
\caption{Spectral energy distribution of the individual a and b spectra, in red and blue, respectively, for the 4 IRS segments. In black the average spectrum for each segment.}
\label{4plots_2001CC21}
\end{figure*}

\twocolumn
\section{Thermal model}
\label{text_model}

\begin{figure}
\centering
\includegraphics[width=0.5\textwidth]{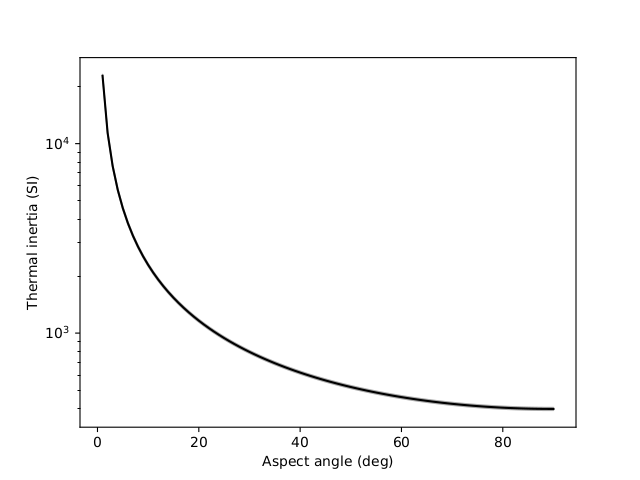}
\caption{Variation of the thermal inertia (SI stands for international system units, which for thermal inertia are J~m$^{-2}$~s$^{-0.5}$~K$^{-1}$, as a function of the aspect angle, from Eq. 2 of \cite{Harris_2016}.}
\label{Tinertia}
\end{figure}
The NEATM model we use \citep{Harris_1998} assumes that the asteroid surface temperature, $T(\theta)$, is controlled only by geometry, namely the sub-solar latitude $\theta$, on the asteroid surface. This temperature is a simple function of the sub-solar temperature $T_\text{SS}$, i.e. $T(\theta)$ = $T_\text{SS} (\cos \theta)^\frac{1}{4}$ for $\theta$ $\leq$ 90$^o$  and T($\theta$)=0 for $\theta$ $>$ 90$^o$, with $T_\text{SS} = \left (S_\odot (1-A) r^{-2}  (\eta \epsilon \sigma)^{-1} \right )^{0.25}$, where $A$ is the bolometric Bond albedo, $r$ the heliocentric distance of the asteroid, $S_\odot$ is the solar constant at 1 au from the Sun, $\sigma$ is the Stefan Boltzmann constant, $\epsilon$ is the emissivity, and $\eta$ is the so-called beaming parameter. In NEATM, an initial guess is made for the geometric albedo $p_V$ and $\eta$ value, so that the diameter $D$ and the $A$-value of the asteroid are calculated from $D (km) = 1329 ~10^{-H/5} \times (p_{V})^{-0.5}$ and $A = q ~p_V$, where $H$ is the absolute magnitude, and $q$ the phase integral \citep{Harris_1998}. Once the temperature distribution is calculated as described above, the infrared flux $f(\lambda)$ is derived from the Planck function multiplied by the emissivity,  integrated over the visible heated portion of the model asteroid shape -- which is controlled only by the phase angle \citep{Harris_1998} -- and multiplied by the factor $D^2 / \Delta^2$, where $\Delta$ is the asteroid-observer distance in au. The model flux is compared to the observed one by calculating a $\chi^2$ figure of merit. A least-square fit is performed by minimizing the $\chi^2$ as a function of the model parameters $D$, $\eta$, and $p_V$ (or $A$), to obtain a NEATM solution and its accuracy in terms of size, beaming parameter, and albedo.

\end{appendix}

\end{document}